\documentclass{article}

\usepackage{microtype}
\usepackage{graphicx}
\usepackage{subcaption}
\usepackage{algorithm}
\usepackage{algorithmic}
\usepackage{booktabs, arydshln} 
\usepackage{multirow}
\usepackage{enumitem}
\usepackage{soul}        
\usepackage[table]{xcolor}
\usepackage{amsmath}
\usepackage{amsfonts}
\usepackage{xspace}
\usepackage[dvipsnames]{xcolor}

\makeatletter
\def\adl@drawiv#1#2#3{%
    \hskip.5\tabcolsep
    \xleaders#3{#2.5\@tempdimb #1{1}#2.5\@tempdimb}%
            #2\z@ plus1fil minus1fil\relax
    \hskip.5\tabcolsep}
\newcommand{\cdashlinelr}[1]{%
\noalign{\vskip\aboverulesep
       \global\let\@dashdrawstore\adl@draw
       \global\let\adl@draw\adl@drawiv}
\cdashline{#1}
\noalign{\global\let\adl@draw\@dashdrawstore
       \vskip\belowrulesep}}
\makeatother

\usepackage{hyperref}
\usepackage{cleveref}

\definecolor{darkgreen}{RGB}{0,140,0}

\definecolor{quoted}{HTML}{0066ff}

\sethlcolor{quoted!10}
\DeclareRobustCommand{\quoteda}[1]{{\sethlcolor{quoted!60}\hl{#1}}}

\DeclareRobustCommand{\quotedc}[1]{{\sethlcolor{quoted!40}\hl{#1}}}

\DeclareRobustCommand{\quotedd}[1]{{\sethlcolor{quoted!75}\hl{#1}}}

\DeclareRobustCommand{\quotede}[1]{{\sethlcolor{quoted!10}\hl{#1}}}
\DeclareRobustCommand{\quotedf}[1]{{\sethlcolor{quoted!5}\hl{#1}}}

\usepackage[accepted]{mlsys2025}

\newcommand{\name}{\texttt{dRAG}\xspace}

\mlsystitlerunning{A Decentralized Retrieval Augmented Generation System with Source Reliabilities Secured on Blockchain}

\begin{document}

\twocolumn[
\mlsystitle{A Decentralized Retrieval Augmented Generation System with Source Reliabilities Secured on Blockchain}



\mlsyssetsymbol{equal}{*}

\begin{mlsysauthorlist}
\mlsysauthor{Yining Lu}{equal,nd}
\mlsysauthor{Wenyi Tang}{equal,nd}
\mlsysauthor{Max Johnson}{nd}
\mlsysauthor{Taeho Jung}{nd}
\mlsysauthor{Meng Jiang}{nd}
\end{mlsysauthorlist}

\mlsysaffiliation{nd}{Department of Computer Science and Engineering, University of Notre Dame, USA}

\mlsyscorrespondingauthor{Yining Lu}{ylu33@nd.edu}
\mlsyscorrespondingauthor{Wenyi Tang}{wtang3@nd.edu}

\mlsyskeywords{Machine Learning, MLSys, decentralized RAG, Blockchain}

\vskip 0.3in

\begin{abstract}
Existing retrieval-augmented generation (RAG) systems typically use a centralized architecture, causing a high cost of data collection, integration, and management, as well as privacy concerns. There is a great need for a decentralized RAG system that enables foundation models to utilize information directly from data owners who maintain full control over their sources. However, decentralization brings a challenge: the numerous independent data sources vary significantly in reliability, which can diminish retrieval accuracy and response quality.
To address this, our decentralized RAG system has a novel reliability scoring mechanism that dynamically evaluates each source based on the quality of responses it contributes to generate and prioritizes high-quality sources during retrieval. To ensure transparency and trust, the scoring process is securely managed through blockchain-based smart contracts, creating verifiable and tamper-proof reliability records without relying on a central authority.
We evaluate our decentralized system with two Llama models (3B and 8B) in two simulated environments where six data sources have different levels of reliability. Our system achieves a +10.7\% performance improvement over its centralized counterpart in the real world-like unreliable data environments. Notably, it approaches the upper-bound performance of centralized systems under ideally reliable data environments. The decentralized infrastructure enables secure and trustworthy scoring management, achieving approximately 56\% marginal cost savings through batched update operations. Our code and system are open-sourced at \href{https://github.com/yining610/Reliable-dRAG}{github.com/yining610/Reliable-dRAG}.
\end{abstract}
] 



\printAffiliationsAndNotice{\mlsysEqualContribution} 

\section{Introduction}
\begin{figure}[ht]
    \centering
    \includegraphics[width=\linewidth]{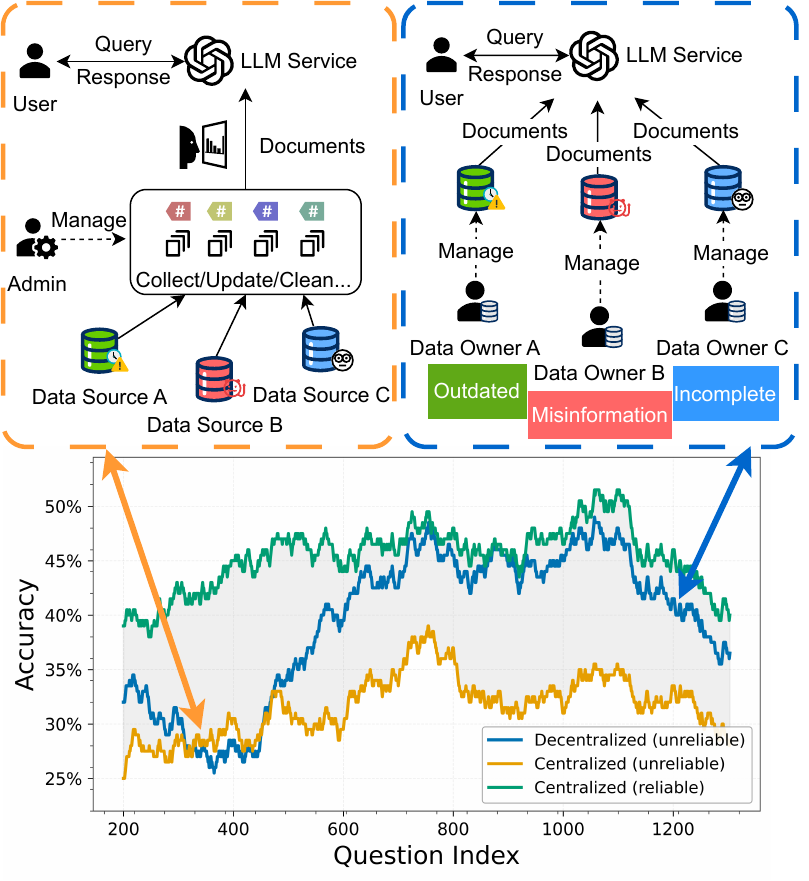}
    \caption{Overview of a centralized RAG system (\textbf{left}) and our \name{} system (\textbf{right}), with their performance comparison (\textbf{bottom}) on \textsc{Llama-3.2-3B-Instruct}. In the unreliable data environments, our \name{} significantly outperforms the centralized system with increasing query exposure, approaching the performance upper-bound from the centralized system achieved in fully reliable data environments.}
    \label{fig: teaser figure}
\end{figure}

Retrieval Augmented Generation (RAG) \citep{NEURIPS2020_6b493230} has been a popular technique for enhancing large language models (LLMs) by providing them access to external knowledge sources during inference. Most existing RAG systems adopt a centralized architecture in which a single administrator manages all data sources, including collection, cleaning, de-duplication, and indexing. 
However, as the number of data sources grows, this centralized architecture encounters challenges in data management costs \citep{gao2024retrievalaugmentedgenerationlargelanguage, barnett2024sevenfailurepointsengineering} and privacy concerns \citep{addison2024cfedragconfidentialfederatedretrievalaugmented}. 
To address these challenges, we build a decentralized RAG system (\textbf{\name{}}) that adopts the similar definition of data providers introduced by \citet{11014986}. Our \name{} system enables LLM services to retrieve documents directly from sources maintained and controlled by individual data owners, giving them the flexibility to decide what information to share and what retrieval policies to implement.

However, building \name{} is not trivial because data sources from different owners often vary notably in quality, making it impractical to treat all sources equally reliable. For example, some data sources may contain \textcolor{ForestGreen}{outdated}, \textcolor{Red}{misinformation}, or \textcolor{NavyBlue}{incomplete} data (as shown in \autoref{fig: teaser figure}). 
The existing decentralized RAG systems \citep{11014986, xu2025distributedretrievalaugmentedgeneration, zhou2025dgragdistributedgraphbasedretrievalaugmented} typically assume all data sources are fully reliable, which is unrealistic in practice.
To overcome this limitation, \name{} has a novel scoring mechanism that dynamically evaluates the reliability of data sources from the quality of responses they help generate. These reliability scores enable \name{} to prioritize high-quality sources during retrieval, thereby improving both retrieval accuracy and response quality in noisy data environments.  
Yet these reliability scores cannot be trusted if managed by a centralized entity, as a centralized administrator could easily manipulate scores to favor certain data sources, creating a single point of failure and compromising system integrity.
So, we leverage blockchain in \name{} to secure the reliability scores and relevant logs. The scoring mechanism is implemented within a smart contract, whose execution and verification are collectively performed by the majority of nodes in the blockchain network.
This ensures that all updates to the scores are executed properly, creating tamper-proof and transparent scoring records that all participants can verify without relying on a central authority.

\name{} can benefit many domains and applications where data cannot be centrally managed, such as regulated enterprises (e.g., healthcare and public sector), cross-institutional consortia, multi-tenant platforms, and open-source communities. \name{} offers reliable retrieval augmentation across independently owned sources through two mechanisms: (1) client-side reliability scoring for response quality, and (2) a blockchain-based reliability management interface ensuring transparency and trustworthiness. 
\name{} can be easily deployed: data providers expose a standards-compliant retrieval endpoint and register with the smart contract registry, while users install our lightweight, open-sourced client library, which connects to the deployed smart contract on-chain for retrieval and reranking.
Furthermore, the continuously updated reliability scores on blockchain provide each data owner with actionable feedback, encouraging proactive monitoring and improvement of their data sources.

To demonstrate the effectiveness of \name{}, we simulate unreliable data environments in which multiple data sources provide Wikipedia documents of varying reliability to answer questions from the Natural Questions dataset \cite{kwiatkowski-etal-2019-natural}.
Specifically, we inject noise into documents by replacing ground-truth tokens with incorrect answers, thereby polluting them to varying levels of reliability. \textbf{An effective \name{} system should demonstrate robustness in such unreliable environments by outperforming its centralized counterpart} (see \textit{Centralized (unreliable)} and \textit{Decentralized (unreliable)} lines in \autoref{fig: teaser figure}) and exhibit consistent growth as more user queries are processed.
Remarkably, our results also show that \textbf{\name{} can even approach the performance of a centralized system with fully reliable data sources} (see \textit{Centralized (reliable)} with no injected noise), an ideal configuration that is difficult and costly to maintain in real-world settings. As previously discussed, centralized systems require administrators to invest substantial effort in data management to ensure data source reliability, whereas \name{} eliminates this burden while achieving comparable performance.
In summary, our contributions are threefold:
\begin{itemize}[leftmargin=*]

\item We introduce a reliability scoring mechanism for decentralized RAG systems that dynamically evaluates source reliability to improve retrieval and answer quality.

\item We build our \name{} on blockchain to establish a decentralized, transparent, and trustworthy scoring management for data source reliability, using smart contracts to ensure full traceability of score updates. We open-source our system to facilitate future research.

\item Through controlled experiments, we demonstrate that \name{} can improve retrieval and response quality over time, outperforming centralized systems under unreliable data environments and achieving performance comparable to those with  ideal, fully reliable data sources.
\end{itemize}

\section{Related Work and Background}

\subsection{Centralized Retrieval Augmented Generation}
Traditional RAG systems rely on centralized architectures in which knowledge bases and retrieval mechanisms are managed under unified control \citep{karpukhin-etal-2020-dense, NEURIPS2020_6b493230, li2022surveyretrievalaugmentedtextgeneration, gao2024retrievalaugmentedgenerationlargelanguage}. This centralized design introduces several fundamental limitations that hinder its scalability and practical deployment.

Centralized RAG architectures face inherent scalability limitations stemming from their single-index design and monolithic control \citep{wang2024mragreinforcinglargelanguage, douze2025faisslibrary}. Even with techniques such as query routing and hierarchical retrieval \citep{xu2025distributedretrievalaugmentedgeneration, helmi2025decentralizingaimemoryshimi}, such systems struggle to efficiently scale as data volumes and domains increase. 
Beyond scalability, centralization leads to data management complexity across diverse data sources \citep{chong2025llmnetdemocratizingllmsasaserviceblockchainbased}, high privacy risks \citep{zeng-etal-2024-good, zeng2025mitigatingprivacyissuesretrievalaugmented}, and governance challenges related to access control and compliance \citep{jayasundara2024ragentretrievalbasedaccesscontrol, zeng-etal-2025-rag}. By contrast, our \name{} inherently mitigates these issues by distributing retrieval across independent data providers for better scalability and data management.

\subsection{Decentralized Retrieval Augmented Generation}
Recent studies have explored decentralized solutions to these limitations through different technical strategies \citep{11014986, yang2025agentnetdecentralizedevolutionarycoordination, zhou2025dgragdistributedgraphbasedretrievalaugmented, chakraborty2025federatedretrievalaugmentedgenerationsystematic}. The External Retrieval Interface framework aims to decentralize the RAG system from a software perspective by decoupling retrieval, augmentation, and generation components through standardized protocols, enabling data providers to maintain full control and keep data at its source \citep{11014986}. 
Another notable approach is federated learning, which enables collaborative training while preserving data locality. For example, FedRAG demonstrates privacy-preserving training of RAG components without raw data exchange \citep{mao2025privacypreservingfederatedembeddinglearning}. De-DSI applies federated learning to train differentiable search indexes without centralized coordination \citep{10.1145/3642970.3655837}. Unlike these approaches, which require complex training and coordinated parameter sharing, our \name{} avoids training overhead by operating entirely at inference time and emphasizes traceability through blockchain logging of source contributions, providing a lightweight yet trustworthy solution for quality control.

Blockchain is considered a natural choice for building decentralized applications, including decentralized RAG systems. 
It helps establish a decentralized, immutable ledger through cryptographic hashing and peer-to-peer consensus mechanisms, ensuring data integrity and transparency without the need for a central authority. 
Smart contracts are Turing-complete programs deployed on the ledger. 
It provides guaranteed execution integrity, as its code and state are verified and validated by the entire decentralized network via the underlying consensus mechanism, as long as the majority of the network is honest. 
The consensus ensures that the contract logic executes exactly as programmed, with the resulting state change being recorded on the tamper-proof blockchain ledger only after achieving network-wide agreement.
This allows the development of robust, transparent, and tamper-proof decentralized applications (dApps) that run on a decentralized network with public verifiability, rather than relying on single/institutional trust.
Existing blockchain-based RAG approaches employ consensus protocols for decentralized knowledge validation. For example, some frameworks introduce domain-specific validator networks (e.g., using distributed hash tables) \citep{10885343} or permissioned blockchains where expert nodes must review and agree on new knowledge before it is integrated \citep{EAndersen2025DRAGAP}. 

Unlike these mechanisms that rely on complex blockchain protocols to coordinate knowledge propagation, our approach uses smart contracts in a more lightweight manner. It leverages smart contracts to dynamically evaluate and prioritize data sources based on their reliability, eliminating the need for complex protocols with specialized validator committees or centralized content reindexing.
Essentially, we provide a transparent, tamper-proof trust layer on top of decentralized retrieval, enabling robust performance even when some sources are unreliable. 

\section{System Architecture}

\subsection{System Components}
\label{subsec: system architecture}

\begin{figure}[ht]
    \centering
    \includegraphics[width=\linewidth]{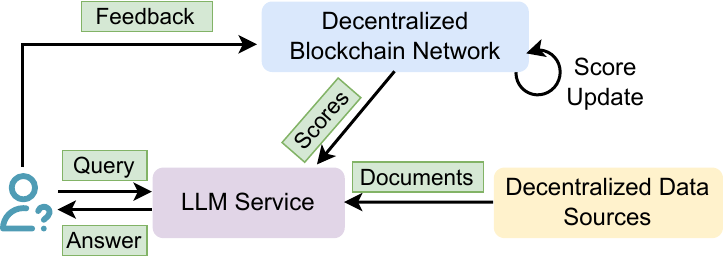}
    \caption{System overview of \name.}
    \label{fig: architecture}
\end{figure}

Our \name{} system can be abstracted into three components: \textit{Decentralized Data Sources}, \textit{LLM Service}, and \textit{Decentralized Blockchain Network} (as shown in \autoref{fig: architecture}).

\noindent \textbf{Decentralized Data Sources.} The decentralized data sources in \name are maintained individually by each data owner, and each may provide a standard-compatible API for the LLM service to query for documents. 

\noindent \textbf{Decentralized Blockchain Network.} The decentralized blockchain network serves as the infrastructure for \name{} to evaluate the reliability of each decentralized data source (detailed in \Cref{sec: score framework}). 
The scoring management is implemented via a smart contract on the Ethereum blockchain, providing trustworthy external evaluation of decentralized data sources. The smart contract, which maintains the scoreboard for all data sources, is executed by a public, peer-to-peer blockchain network. The integrity of its execution is guaranteed by the decentralized nature of blockchain.  

\noindent \textbf{LLM Service.} The LLM service responds to user queries, interacting with other system components, including the smart contract on the blockchain, and the decentralized data sources to generate the answer. Notably, 
multiple LLM services can share the same \name{} infrastructure to serve different users by connecting and providing feedback to the same smart contract on-chain.

\subsection{System Workflow}
\autoref{fig: architecture} presents the system overview of \name. Additional implementation details are provided in \Cref{sec: blockchain}. We assume the LLM Service has direct access to individual data sources, allowing it to retrieve the most relevant documents through standard-compliant interfaces. 

Similar to other RAG systems, users submit queries to the LLM service to obtain answers from \name.
The LLM service will first sample a subset of data sources based on the reliability evaluation from \Cref{subsec: importance evaluation}, and route the query to them. 
The sampled data sources then use the locally hosted retriever to find the documents most relevant to the query and return them to the LLM service.
The data source sampling process can be customized to meet the query and user needs. 
In \name, we present an effective sampling process based on a parameter that quantifies the usefulness of each data source in improving the LLM performance, as described in \Cref{subsec: reliability update}. 
After retrieving the documents from data sources, the documents will be reranked and concatenated with the query before being fed to the LLM. 
The reranking process incorporates reliability scores to identify the most reliable documents for answering the query. The LLM then generates an answer, and the user provides feedback on it to update the reliability scores on the blockchain. While our system accepts direct user feedback, we also support automatic evaluation via exact matching to ground-truth answers (if available).
We provide a detailed discussion of reliability scoring, data source sampling, and reranking in the next section.

\section{Evaluating Data Sources Reliability} 
\label{sec: score framework}

\begin{algorithm}[ht]
    \small
    \caption{Reliability-Guided RAG}
    \label{alg: reliability}
    \begin{algorithmic}[1]
      \STATE \textbf{Inputs:} user query $q$; data sources $\mathcal{S}=\{s_i\}$
      \STATE \textbf{Hyperparameters:} number of retrievers $N$; per-source fetch $M$; reranked top-$K$
      \STATE \textbf{Initialization:} usefulness $U_i$ and reliability $R_i$ for source $s_i$
      \FOR{each query $q$}
        \STATE Sample $N$ sources $\{s_i\}$ proportionally to $U_i$
        \STATE Retrieve $M$ documents from each selected source $\mathcal{X}$ with size $M\times N$
        \STATE Rerank $\mathcal{X}$ with \autoref{eq: rerank}; keep top-$K$ documents $\mathcal{X}_{1:K}$
        \STATE Generate response $y$ using $\mathcal{X}_{1:K}$
        \STATE For each document $d\!\in\!\mathcal{X}_{1:K}$ and sentence $x\!\in\!d$, compute $f(x)$ via either \autoref{eq: unsupervised} or \autoref{eq: supervised}
        \STATE Compute document score: $f(d)\!\leftarrow\!\mathrm{Agg}_{x\in d} f(x)$
        \STATE Update $U_i, R_i$ for each source $s_i$ with documents in $\mathcal{X}_{1:K}$ using $f(d)$ following \autoref{eq: reliability update}
      \ENDFOR
    \end{algorithmic}
\end{algorithm}

We introduce how \name{} evaluates (\S\ref{subsec: importance evaluation}) and updates (\S\ref{subsec: reliability update}) data source reliability to improve retrieval and response quality. We present the whole procedure in \Cref{alg: reliability}.

\subsection{Sentence Importance Evaluation}
\label{subsec: importance evaluation}
The utility of retrieved documents varies across different information needs and temporal contexts \citep{qian-etal-2024-timer4, uddin-etal-2025-unseentimeqa}. For instance, a Wikipedia article documenting the 2020 U.S. presidential election provides limited value when addressing queries about the 2024 election. We consider a data source \textit{\textbf{reliable}} if it consistently provides informative content that improves LLM generation across a diverse range of user queries. To quantify this notion of reliability, we first compute the importance of sentences from the retrieved document to LLM generation, which can be evaluated under two scenarios.

\paragraph{Ground-truth answers are unavailable.} We estimate the importance of each sentence $x$ in the retrieved document $d$ using Monte-Carlo Shapley (\textbf{MC-Shapley}) values \citep{goldshmidt2024tokenshapinterpretinglargelanguage}. Formally, the reliability score is defined as the expected marginal contribution of $x$ to the model output across sampled subsets of sentences:
\begin{equation}
        f_{\text{shapley}}(x) = \mathbb{E}_{s \subseteq d \setminus \{x\}} \big[ F(s \cup \{x\}) - F(s) \big],
        \label{eq: unsupervised}
\end{equation}
where $F(\cdot)$ is the utility function, which in our setting is computed as the cosine similarity between the baseline response (generated using the full retrieved document) and the response conditioned on the sampled subset $s$ or $s \cup \{x\}$. This formulation quantifies the marginal contribution of each sentence to the model’s output, featuring \textit{greater generalizability but with increased computational overhead.}
    
\paragraph{Ground-truth answers are available.} We adopt an information-theoretic formulation in which each sentence is treated as a potential rationale. The informativeness of a rationale, denoted as $f_{\text{info}}(x)$, is defined as the conditional $\mathcal{V}$-information \citep{hewitt-etal-2021-conditional} capturing the reduction in model predictive uncertainty conditioning on the rationale:
\begin{equation}
    f_{\text{info}}(x) = H_{\mathcal{V}}(y^\ast \mid q) - H(y^\ast \mid q \;\cup\; x),
    \label{eq: supervised}
\end{equation}
where $y^\ast$ and $q$ are the ground-truth answer and query. We train evaluators following \textbf{RORA} \citep{jiang-etal-2024-rora} to estimate $H(\cdot)$ as a multivariable predictive $\mathcal{V}$-entropy. This approach \textit{improves computational efficiency but reduces generalizability to out-of-domain sentences}.

\subsection{Data Source Reliability Update}
\label{subsec: reliability update}
Given sentence-level importance scores $\{f(x) : x \in d\}$ for a document $d$ retrieved from data source $s_i$, we aggregate these scores to obtain a document-level reliability estimate $f(d)$. The aggregation strategy depends on the score scale and the corresponding evaluation method: for Shapley-based scores, we compute the arithmetic mean $f(d) = \frac{1}{|d|} \sum_{x \in d} f_{\text{shapley}}(x)$, 
while for entropy-based scores, we take the maximum $f(d) = \max_{x \in d} f_{\text{info}}(x)$.

We maintain a cumulative reliability score $R_i$ for each data source $i$, which is updated based on user feedback regarding the correctness of the generated response $y$. The update rule is straightforward: for document $d$ belonging to source $s_i$,
\begin{equation}
    R_i \leftarrow \begin{cases}
        R_i + f(d), & \text{if } y \text{ is correct} \\
        R_i - f(d), & \text{if } y \text{ is incorrect}
    \end{cases}
    \label{eq: reliability update}
\end{equation}

This additive update scheme allows the system to accumulate evidence about source reliability over multiple interactions, rewarding sources that contribute to correct answers while penalizing those with errors. 

While reliability measures source accuracy, our experiments show that LLMs occasionally generate responses without using the retrieved documents. In such cases, a source may be inherently reliable yet fail to contribute usefully to the generation. To address this gap, we introduce a complementary usefulness score $U_i$ that quantifies the extent to which documents from source $i$ are actually used by the LLM to generate responses. We evaluate whether generated responses can be grounded in the retrieved documents and update $U_i$ accordingly: (1) If the response is \textbf{not grounded}, we penalize the usefulness score as $U_i \leftarrow U_i - f(d)$, while leaving $R_i$ unchanged. (2) If the response \textbf{is grounded}, we reward usefulness $U_i \leftarrow U_i + f(d)$ and update reliability score $R_i$ following \autoref{eq: reliability update}. We provide empirical justification for $U_i$ in \Cref{subsec: sensitivity analysis}.

\paragraph{Reliability-Aware Retrieval and Reranking.}
We incorporate these scores at different RAG stages to optimize retrieval quality. Following the standard RAG pipeline \citep{karpukhin-etal-2020-dense, li-etal-2025-enhancing-retrieval}, we employ a dense retrieval system implemented with FAISS \citep{douze2025faisslibrary} for initial candidate selection, followed by a neural reranker for final document ordering.  Specifically, at the retrieval stage, we use usefulness scores $U_i$ to sample $N$ dense retrievers and their corresponding data sources. At the reranking stage, we leverage source reliability $R_i$ to refine the final document ordering. For a candidate document retrieved from data source $s_i$, its final reranking score is:
\begin{equation}
    \text{score}_{\text{final}} = (1 - \alpha)\cdot \text{score}_{\text{rerank}} + \alpha \cdot \sigma(R_i)
    \label{eq: rerank}
\end{equation}
where $\text{score}_{\text{rerank}}$ denotes the original reranking score and $\sigma(\cdot)$ is a normalization function (implemented as min–max normalization in our case). The hyperparameter $\alpha \in [0,1]$ controls the influence of reliability on the final ranking.

\section{On-Chain Reliability Management System with Smart Contract}
\label{sec: blockchain}
\begin{figure}[ht]
    \centering
    \includegraphics[width=\linewidth]{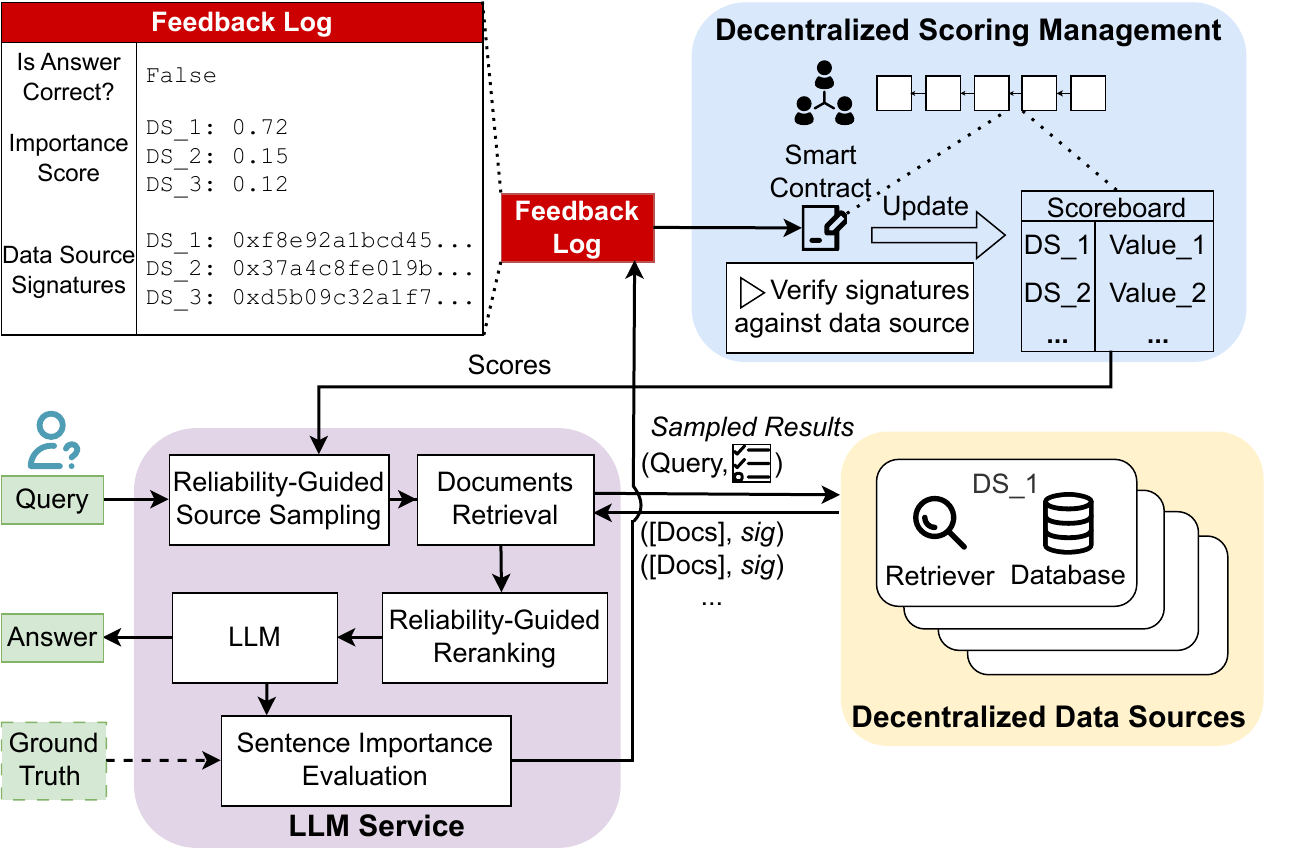}
    \caption{Detailed Workflow of \name, with a feedback log to update the reliability scores. For brevity, state information linking the update to a genuine query is omitted in the example. Ground truth is optional for sentence importance evaluations; if it is unavailable, user feedback (\texttt{True/False}) will be used to update the reliability scores.}
    \label{fig: detail workflow}
\end{figure}

\begin{algorithm}[ht]
\caption{Trustworthy Reliability Update with Signature}
\label{alg:trustworthy_update}
\begin{algorithmic}[1]
\STATE \textbf{Inputs:} State digest $\mathit{state}$, signatures $\mathit{sig}[1..n]$, source IDs $\mathit{Source}[1..n]$,
        updated reliability scores $\mathit{R}[1..n]$, updated usefulness scores $\mathit{U}[1..n]$, feedback info $\mathit{info}$

\STATE $t \gets \text{current\_timestamp()}$
\IF{lengths of $\mathit{sig}$, $\mathit{Source}$, $\mathit{R}$, $\mathit{U}$ differ}
    \STATE \textbf{abort} with \texttt{InvalidUpdateCount}
\ENDIF

\FOR{$i = 1$ to $n$}
    \IF{score record for $\mathit{Source}[i]$ does not exist}
        \STATE \textbf{abort} with \texttt{DataSourceNotExists}
    \ENDIF
    \STATE $\mathit{pk}_i \gets \text{scoreRecords}[\mathit{Source}[i]].\text{sourceAddress}$
    \IF{$\neg\,\text{ECDSA.verify}(\mathit{state}, \mathit{sig}[i], \mathit{pk}_i)$}
        \STATE \textbf{abort} with \texttt{InvalidSignature}
    \ENDIF
    \STATE Update $\text{scoreRecords}[\mathit{Source}[i]]$: $\text{reliabilityScore} \gets \mathit{R}[i]$, $\text{usefulnessScore} \gets \mathit{U}[i]$, $\text{timestamp} \gets t$
    \STATE Emit \texttt{ScoreRecordUpdated} event for $\mathit{Source}[i]$ with new scores, $t$, and $\mathit{info}$
\ENDFOR
\end{algorithmic}
\end{algorithm}

In this section, we describe the detailed system implementation of \name{} with the reliability evaluation from \Cref{sec: score framework}.

\subsection{On-Chain Trustworthy Reliability Management with Smart Contract}
\label{subsec: on-chain reliability}

By distributing the knowledge base and eliminating the need for a centralized database manager, \name requires decentralized management of source scores.
Since the reliability scores are used to evaluate the trustworthiness of data sources, they must be managed on a public bulletin board where no single party can tamper with them. Additionally, the updating of these scores should allow public auditing to ensure transparency. 
This design makes the scoring mechanism in \name more trustworthy and transparent.

\name utilizes blockchain as an infrastructure for decentralized score management and auditable logging.
A smart contract is deployed on the blockchain to facilitate the implementation of basic scoring functions, including initialization, retrieval, and update. Every update of the scores will leave a log on-chain, with traceable information that showcases the reason for the modification.
The blockchain serves as a public bulletin board to keep the scores, and the smart contract provides the interface to manage and update the scores with decentralization-based execution integrity. 

\noindent\textbf{Score Initialization.}
In \name, every data source owner will be registered as an account with an ECDSA (Elliptic Curve Digital Signature Algorithm) compatible public/secret key pair $(pk, sk)$ on the blockchain. 
Every owner of the data source needs to use the deployed \name contract to initialize a scoring record for their data source on the blockchain. The contract will create corresponding score records for each data source, each with an initial score, which will be stored as a state variable on-chain and be publicly accessible. 

\noindent \textbf{Reliability Score Update.}
After obtaining responses from the LLM in \name, users can provide feedback on responses along with ground-truth answers, and update the reliability based on their contribution to the reliability records on the blockchain via a smart contract. 
The LLM service will first perform a sentence-level importance evaluation (described in \Cref{sec: score framework}) and calculate score changes based on the evaluation. 
Then, the LLM service needs to include the evaluation in a feedback log transaction and submit it to the blockchain network so the deployed smart contract can update the scores for each source. To prevent arbitrary updates to reliability records, the smart contract will perform basic verification of the feedback log to ensure that updates can be traced back to a legitimate query sent to the data sources.

\noindent \textbf{Trustworthy Reliability Update with Signature.}
\Cref{alg:trustworthy_update} presents the pseudocode for the signature verification and reliability score update logic of our smart contract. We also provide a deployed smart contract instance of this algorithm at the public testnet Sepolia at \href{https://2at6kd.short.gy/4R9Tbm}{2at6kd.short.gy/4R9Tbm}.

When the user's query is broadcast to the sampled data sources to retrieve potentially relevant documents, the LLM service also generates $state$ information and attaches it to the query.
The $state$ is a digital digest (secure hash) that includes the query and the reliability scores of all sampled data sources, which can be viewed as a summary binding the current query to the current state of all sampled data source scores.
The sampled data sources can obtain the same information (the query and the reliability scores of all data sources) to generate and verify if the $state$ matches their current view. 
When returning the retrieved documents to the LLM service, each sampled data source must create a digital signature $sig$ using its private key $sk$, indicating its consent to acknowledge and confirm the $state$. This signature will be considered as consent to allow the LLM service to update the reliability score of the data sources.  
The LLM service creates the feedback log, and the signatures obtained from each data source will be included as part of the feedback log transaction updating the scores, as shown in \autoref{fig: detail workflow}. 

The smart contract will retrieve all relevant reliability records associated with the owners' public keys and perform batch verification of all signatures against the information included in the transaction, including the query prompt, the importance evaluation summary, and the reported scoring state hash from the data sources.  
The batched verification of the above information ensures that the feedback log transaction, which updates the scores, can be traced back to a specific user query.

For further security and tracing purposes, like preventing replay attacks by reusing the signature to maliciously update the data source reliability scores, one may consider leaving a log recording the query and sampled data source list (marked as ``unused'') on-chain before the LLM service broadcasts the query and retrieves the documents from data sources. When the data sources create the signature, the $state$ should also include the query log. The first feedback on the query marks the query log as ``used'' and invalidates all subsequent updates that attempt to exploit the same query.

\subsection{Auditing Reliability Updates}
Whenever a score update occurs for any data source, the deployed smart contract will emit a \texttt{ScoreRecordUpdated} event, including the necessary information to trace the update back to the original query included in the feedback log (as shown in line 14 of \Cref{alg:trustworthy_update}). 
The event data will be part of the blockchain but not directly accessible to the smart contract, helping reduce the network's computational overhead. The event data can be queried and retrieved by any blockchain gateway node, linked to the specific transaction in the blockchain, allowing anyone to audit the entire history of data source reliability updates and ensure the update can be traced back to a genuine query. 
This information can be used to reproduce the query for further investigation.

Additionally, our system can also be extended to support asynchronous feedback and score updates from the users. This could be useful in cases where ground truth for certain queries is not immediately available, allowing users to provide evaluations and update scores at a later time.
Specifically, users still receive signatures from selected data sources and obtain responses from the LLM. The sentence-level importance evaluation will be conducted without ground truth using \autoref{eq: unsupervised}. Users can submit feedback with a non-determined importance report to the blockchain. Once ground truth becomes available, users can invoke the smart contract to report it on-chain and then update the reliability scores accordingly.

\section{Experiments and Results}
\begin{figure}[t]
    \centering
    \includegraphics[width=\linewidth]{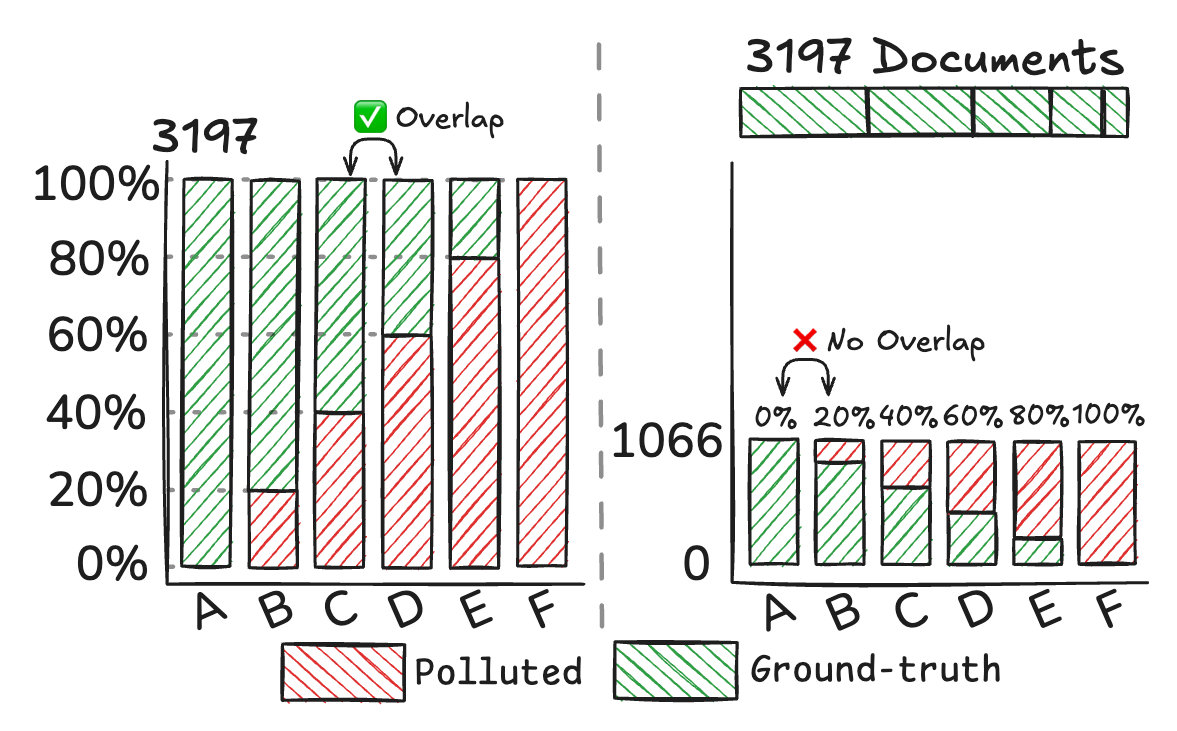}
    \caption{Statistics of two simulated unreliable data environments. \textbf{Left}: Token-level pollution shows the percentage of ground-truth tokens replaced with random tokens across six data sources (\textit{A-F}), where different sources contain overlapping ground-truth documents. \textbf{Right}: Document-level pollution shows how 3197 ground-truth documents are divided into six sources (\textit{A-F}) and mixed with polluted documents to create varying reliability levels, with no document overlap between sources.}
    \label{fig: data stat}
\end{figure}

To demonstrate the effectiveness of \name{}, we evaluate it across two environmental configurations, each containing 6 data sources with varying pollution levels $p \in \{0\%, 20\%, \cdots, 100\%\}$, where $p=0\%$ indicates no pollution (i.e., the data source provides ground-truth documents for the original Natural Questions dataset \citep{kwiatkowski-etal-2019-natural}), and higher values of $p$ indicate increasingly unreliable documents within the data source. For building these two environments, we (1) inject noise into the documents by replacing ground-truth tokens with randomly sampled tokens based on the given pollution level (\textbf{token-level pollution}); (2) partition the original dataset into 6 disjoint data sources, where each source is populated with randomly sampled polluted documents, such that all sources maintain equal length and achieve their specified pollution levels without overlap (\textbf{document-level pollution}).\footnote{We use the words ``polluted'' and ``unreliable'' interchangeably hereafter.} \autoref{fig: data stat} shows statistics of these two data environments.

For polluted data environments, an effective reliability evaluation should successfully identify and prioritize the least polluted (i.e., most reliable) data sources throughout the process, and therefore a well-designed \textbf{\name{} system should outperform its centralized counterpart under polluted environments, while ideally approaching the performance upper bound set by centralized systems in unpolluted conditions.}

We evaluate our \name{} system on \textsc{Llama3.2-3B-Instruct} and \textsc{Llama-3.1-8B-Instruct} \citep{grattafiori2024llama3herdmodels}. We train our evaluators using invariant learning with T5-base \citep{2020t5} to compute \autoref{eq: supervised}, following the RORA approach \citep{jiang-etal-2024-rora}. We use the following hyperparameters in our main experiments (for results shown in \autoref{table: main results}): reliability weight $\alpha=0.5$ for token-level pollution environments and $\alpha=0.2$ for document-level pollution environments.\footnote{The lower $\alpha$ in document-level settings accounts for the non-overlapping document distribution, where a single high $\alpha$ value would bias retrieval toward completely irrelevant data sources. Further analysis is provided in \Cref{subsec: ce vs de}.} We set the sampling number to $N=5$, top-$K=2$, and per-source fetch $M=3$. For LLM inference, we consistently set the temperature to $0$ for better reproducibility. We initialize both usefulness and reliability scores to 10.

We deploy the smart contract used in \name{} on the Sepolia network, a public Ethereum testnet, together with an easy-to-use package tool to test and evaluate the core functions of \name. We provide a detailed cost evaluation for maintaining the \name infrastructure in \Cref{subsec: system cost}.

\subsection{Centralized versus Decentralized RAG System}
\label{subsec: ce vs de}

\begin{figure*}[t]
    \centering
    \begin{subfigure}[t]{0.33\textwidth}
        \centering
        \includegraphics[width=\linewidth]{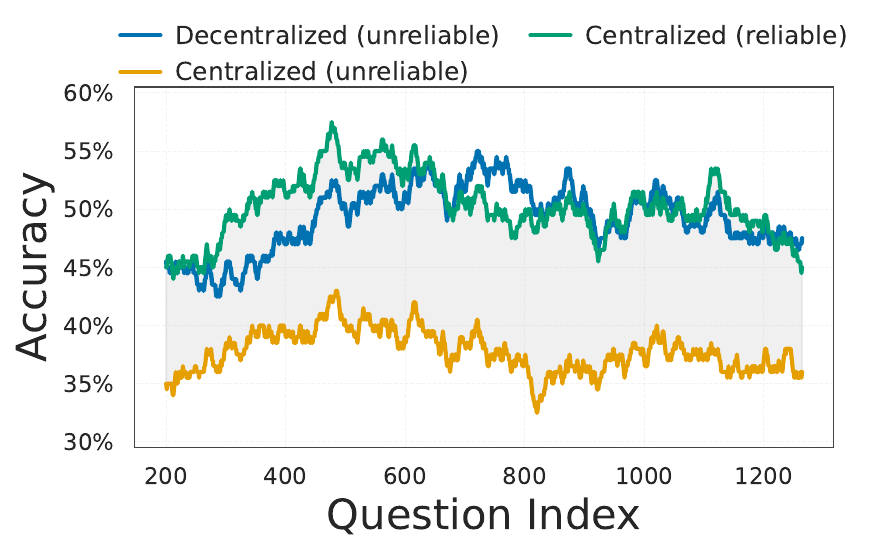}
        \caption{System Performance Comparison}
        \label{fig: ce versus de}
    \end{subfigure}\hfill
    \begin{subfigure}[t]{0.33\textwidth}
        \centering
        \includegraphics[width=\linewidth]{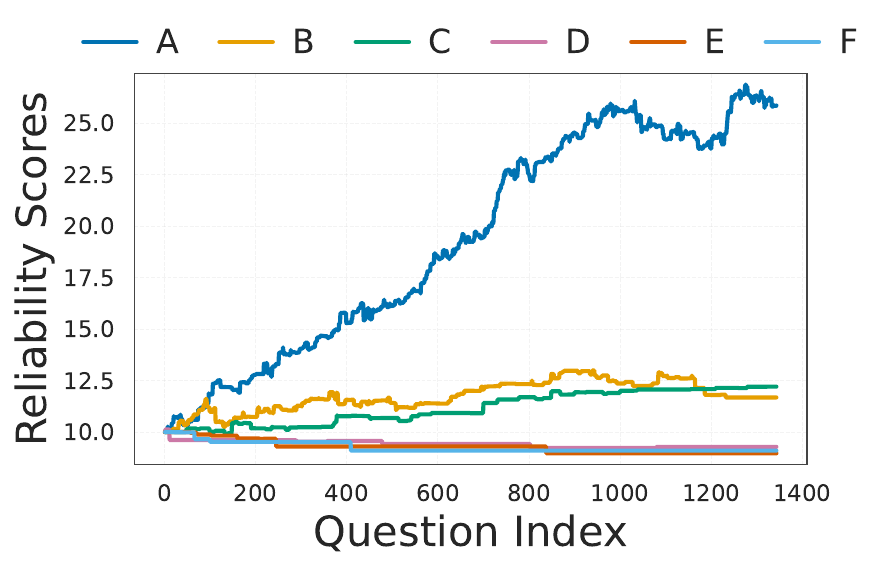}
        \caption{Data Source Reliability}
        \label{fig: reliability score}
    \end{subfigure}\hfill
    \begin{subfigure}[t]{0.33\textwidth}
        \centering
        \includegraphics[width=\linewidth]{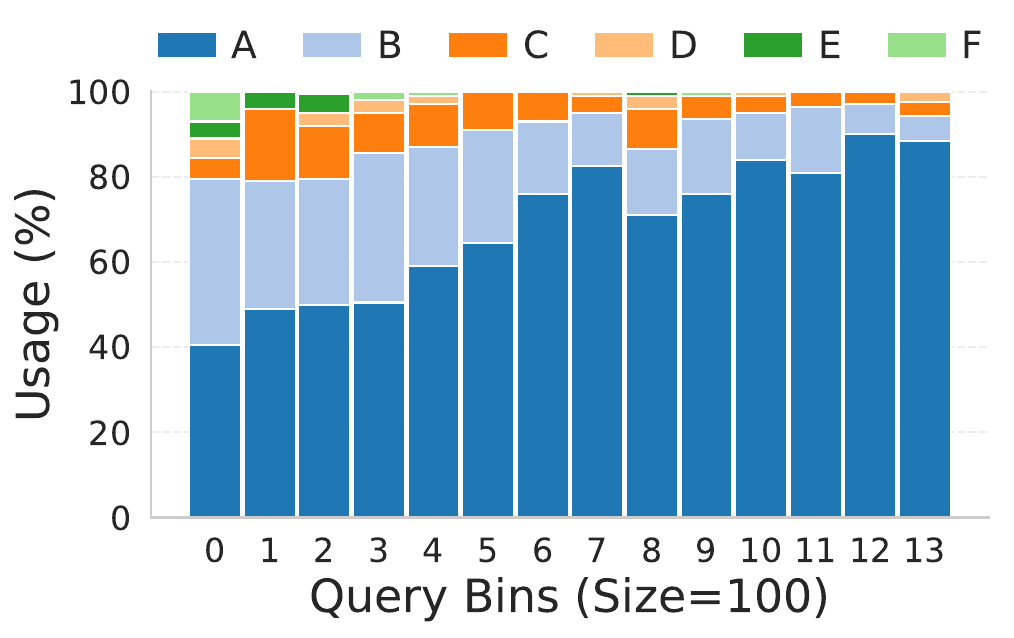}
        \caption{Data Source Usage}
        \label{fig: usage}
    \end{subfigure}
    
    \caption{\textbf{(a)} Performance comparison between \name{} and centralized RAG systems on \textsc{Llama-3.1-8B-Instruct} across reliable and unreliable data environments; \textbf{(b)} Evolution of reliability score $R_i$ (\autoref{eq: reliability update}) throughout the query process; \textbf{(c)} Percentage distribution of data sources (\textit{A-F}) across sequential query bins of 100 queries each.}
\end{figure*}

\begin{table*}[ht]
\small
\centering
\setlength{\tabcolsep}{6pt}
\renewcommand{\arraystretch}{1.15}
\begin{tabular}{@{}c c c cc c@{}}
\toprule
\multirow{2}{*}{\textbf{Models}} &
\multirow{2}{*}{\textbf{Pollution Strategy}} &
\multirow{2}{*}{\shortstack{\textbf{Centralized}\\\textbf{(Unreliable)}}} &
\multicolumn{2}{c}{\textbf{\name{} (Unreliable)}} &
\multirow{2}{*}{\shortstack{\textbf{Centralized}\\\textbf{(Reliable)}}} \\
\cmidrule(lr){4-5}
 & & & \textbf{MC-Shapley} & \textbf{RORA} & \\
\midrule
\multirow{2}{*}{\textsc{Llama-3.2-3B-Instruct}}
 & Token-level    & 30.56 & 42.69 & 49.90 & \multirow{2}{*}{44.24} \\
 & Document-level & 28.74 & 30.60 & 29.90 & \\
\midrule
\multirow{2}{*}{\textsc{Llama-3.1-8B-Instruct}}
 & Token-level    & 37.37 & 46.58 & 43.30 & \multirow{2}{*}{49.03} \\
 & Document-level & 34.38 & 37.37 & 37.24 & \\
\bottomrule
\end{tabular}
\caption{Average accuracy of centralized and decentralized (\name) systems on reliable and unreliable data environments. For \name, performance is reported after 500 warmup queries to ensure reliability score convergence. Reliability scores are updated using sentence-level importance estimation via MC-Shapley \citep{goldshmidt2024tokenshapinterpretinglargelanguage} or RORA \citep{jiang-etal-2024-rora} methods. Unreliable data was polluted using token-level or document-level strategies.}
\label{table: main results}
\end{table*}

\paragraph{\name{} outperforms centralized RAG system in unreliable data environments.} \autoref{fig: ce versus de} presents a performance comparison between centralized and decentralized RAG systems using \textsc{Llama-3.1-8B-Instruct} on token-level polluted data (left in \autoref{fig: data stat}). Combined with the results shown in \autoref{fig: teaser figure} using \textsc{Llama-3.2-3B-Instruct}, these findings demonstrate that \name{} exhibits self-improvement capabilities by progressively learning to prioritize the most reliable data sources across user queries.
The underlying mechanism for this improvement is illustrated in \autoref{fig: reliability score}. Initially, all sources start with identical reliability scores. However, data source \textit{A} consistently receives positive feedback as its high-quality documents are retrieved and contribute to generating correct answers. As a result, its reliability score increases while other sources remain largely unchanged or slightly decline, making \textit{A} more likely to be retrieved in subsequent queries.
This mechanism explains a key behavioral pattern observed in \name{} when deployed in unreliable data environments: the system initially performs at a lower-bound level, equivalent to a centralized RAG system using polluted data, but steadily improves over time as reliable sources accumulate positive feedback and begin to dominate retrieval decisions.

To further validate these improvements, we analyze the usage distribution of each data source across queries. \autoref{fig: usage} confirms that as \name{} digests more queries (organized into bins of 100), it learns to prioritize more reliable sources. Specifically, the usage percentage of data source \textit{A} increases substantially from approximately 40\% in the first query bin to over 80\% in later bins (bins 6-13). This increase is accompanied by a corresponding decrease in the utilization of less reliable data sources \textit{B}, \textit{C}, and \textit{D}, whose combined usage drops from roughly 60\% to under 20\%. These results highlight that \name{} effectively learns to distinguish data quality and progressively favors the least polluted sources, thereby improving retrieval performance over time.

\paragraph{\name{} achieves performance comparable to centralized systems running in ideal, reliable data environments.} \autoref{table: main results} presents the average accuracy of centralized and decentralized systems across various models, scoring methods, and data environments. Our findings demonstrate that \name{} consistently outperforms the centralized system under unreliable data environments across all experimental configurations. 
More remarkably, under conditions of token-level pollution, \name{} achieves performance levels comparable to those of centralized systems operating on fully reliable data sources. This aligns with the intuition that prioritizing high-quality data sources during retrieval fundamentally enhances performance, with the upper bound constrained by the model's inherent capacity to process fully reliable data.

However, we observe a notable performance disparity between the two pollution strategies. Specifically, unreliable environments constructed using document-level pollution yield lower accuracy compared to those using token-level pollution. This difference stems from different document distributions yielded by two strategies (\autoref{fig: data stat}). Document-level pollution creates data sources that cover only a subset of the original documents (1066 out of 3197 to answer queries). 
Consequently, when \name{} promotes the most reliable sources (e.g., \textit{A}) during reranking, it may inadvertently favor irrelevant documents from these sources simply based on their high reliability scores. In contrast, token-level pollution creates a more balanced data environment where all data sources maintain identical document coverage to the original corpus, with reliability varying at the granular token level rather than through selective document inclusion. We leave the challenge of handling these incomplete and unreliable data sources to future work.

\subsection{Sensitivity Analysis and Score Validation}
\label{subsec: sensitivity analysis}

\begin{figure*}[t]
    \centering
    \begin{subfigure}[t]{0.33\textwidth}
        \centering
        \includegraphics[width=\linewidth]{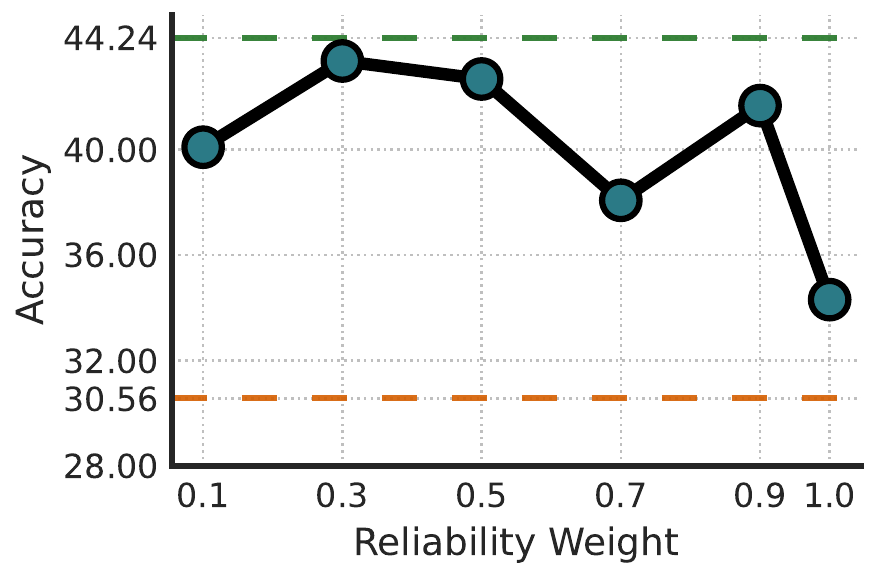}
        \caption{Reweight Weight Sensitivity}
        \label{fig: reliability sensitivity}
    \end{subfigure}\hfill
    \begin{subfigure}[t]{0.33\textwidth}
        \centering
        \includegraphics[width=\linewidth]{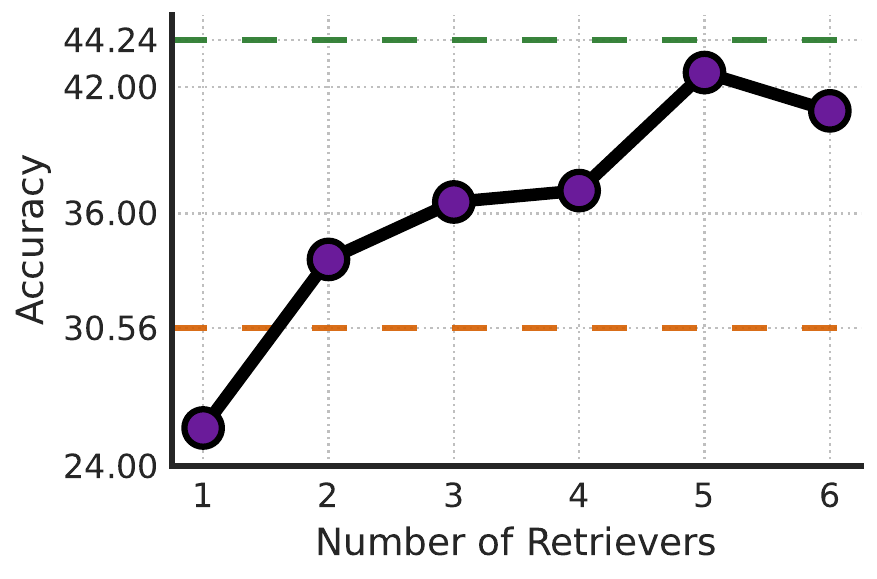}
        \caption{Retriever Number Sensitivity}
        \label{fig: retriever sensitivity}
    \end{subfigure}\hfill
    \begin{subfigure}[t]{0.33\textwidth}
        \centering
        \includegraphics[width=\linewidth]{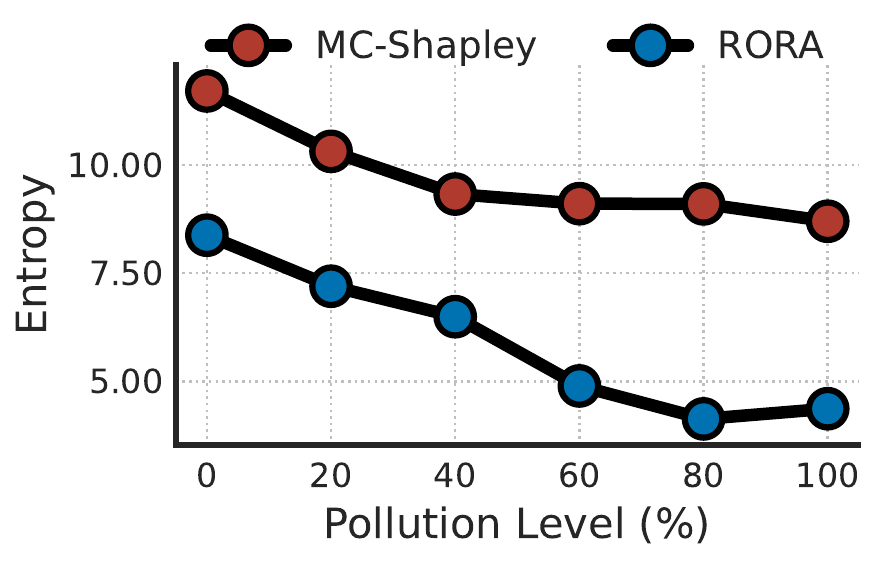}
        \caption{Entropy of Polluted Data Sources}
        \label{fig: entropy}
    \end{subfigure}
    
    \caption{\textbf{(a-b):} Sensitivity test results of \name{} on reliability weight $\alpha$ and number of retrievers $N$. Results are obtained from a token-level polluted environment where \name{} uses the MC-Shapley scoring method with \textsc{Llama-3.2-3B-Instruct}. The \textcolor{orange}{\textemdash} indicates the baseline performance of a centralized system on unreliable data (30.56, from \autoref{table: main results}), and the \textcolor{darkgreen}{\textemdash} indicates the upper-bound performance of a centralized system on reliable data (44.24). \textbf{(c):} Entropy in sentence importance estimation across different pollution levels.}
\end{figure*}

\begin{table*}[ht]
    \centering
    \small
    \renewcommand{\arraystretch}{1.15}

    \begin{tabular}{p{420pt}c}
        \toprule
        \textbf{\textit{Question:}} \textit{Who got the first Nobel prize in physics?} & \textbf{Score}\\
        \midrule

        \multirow{4}{420pt}{%
          \textsc{\textbf{MC-Shapley}}:\,
          \quoteda{The first Nobel Prize in Physics was awarded in 1901 to Wilhelm Conrad Röntgen, of Germany, who received 150,782 SEK, which is equal to 7,731,004 SEK in December 2007.}
          \quotedc{John Bardeen is the only laureate to win the prize twice—in 1956 and 1972.}
          \quotedc{Maria Sklodowska-Curie also won two Nobel Prizes, for physics in 1903 and chemistry in 1911.}\par
        } & 0.3626\\
        & 0\\
        & 0\\
        & \\
        \cdashlinelr{1-2}

        \multirow{4}{420pt}{%
          \textsc{\textbf{RORA}}:\,
          \quotedd{The first Nobel Prize in Physics was awarded in 1901 to Wilhelm Conrad Röntgen, of Germany, who received 150,782 SEK, which is equal to 7,731,004 SEK in December 2007.}
          \quotede{John Bardeen is the only laureate to win the prize twice—in 1956 and 1972.}
          \quotedf{Maria Sklodowska-Curie also won two Nobel Prizes, for physics in 1903 and chemistry in 1911.}\par
        } & 1.0806\\
        & -1.2246\\
        & -2.6737\\
        & \\

        \bottomrule
    \end{tabular}
    \caption{Example sentence-level scores for the first three sentences of the retrieved document, computed by MC-Shapley (using \textsc{Llama-3.2-3B-Instruct}) and RORA for the given question. The scores in the right column correspond to each sentence. The ground-truth answer is ``\textbf{Wilhelm Conrad Röntgen}''. \quoteda{Darker highlighting} denotes sentences assigned higher scores by each method.}
    \label{table: heatmap example}
\end{table*}

Considering that \name{} uses reliability weights $\alpha$ for balancing rerank and reliability scores for each data source, and uses the number of retrievers $N$ for sampling from the usefulness score, we study \name{}'s sensitivity towards the choice of these two hyperparameters.

\paragraph{Reliability weight $\alpha$.} We test \name{} using MC-Shapley across six reliability weight values on \textsc{Llama-3.2-3B-Instruct} under token-level polluted data, where a higher reliability weight results in more aggressive reliability-driven filtering.
Results shown in \autoref{fig: reliability sensitivity} indicate that increasing the weight does
induce stronger intervention (e.g., from 0.1 to 0.5), making the \name{} retrieves documents from less polluted data sources and improves its performance towards the upper bound. However, exceedingly large reliability weights, such as 1.0, can cause the \name{} to over-rely on reliability scores while overlooking the semantic relevance captured by neural rerankers, resulting in accuracy decline to approximately 34\%. 
Notably, across all tested reliability weights, \name{} consistently outperforms the centralized system under identical conditions. 
This further demonstrates \name{}'s robustness in unreliable and noisy data environments, because accidental retrievals from highly polluted sources can be overridden by documents from more reliable sources, which typically receive higher reranking scores according to \autoref{eq: rerank} and are consequently positioned later in the context window where they can receive more attention from LLMs.

\paragraph{Number of retrievers $N$.}
The number of retrievers $N$ determines how many data sources are sampled based on their usefulness scores. A higher value of $N$ forces \name{} to retrieve documents from more data sources that have historically contributed meaningfully to LLM generation (see the usefulness definition in \Cref{subsec: reliability update}). We evaluate $N$ ranging from a single data source to all six available sources, and the results are shown in \autoref{fig: retriever sensitivity}. Clearly, increasing the number of retrievers consistently improves accuracy, elevating performance from below the lower bound to nearly reaching the upper bound. 
This demonstrates that data sources with higher reliability scores are not necessarily the most useful ones for LLM generation. By retrieving documents from multiple sources, \name{} learns from diverse information across different reliability levels, enabling more accurate and nuanced updates to the reliability scores.
We also notice a performance drop when we set $N=6$, where all data sources are included and usefulness-based sampling is practically disabled. Therefore, this drop can be interpreted as an ablation result, highlighting the practical importance of usefulness-based sampling for \name{}.

\paragraph{Sentence importance score validation.} We validate that importance score estimation methods, MC-Shapley \citep{goldshmidt2024tokenshapinterpretinglargelanguage} in \autoref{eq: unsupervised} and RORA \citep{jiang-etal-2024-rora} in \autoref{eq: supervised} can effectively capture the contribution of each retrieved sentence to LLM generation. While these methods have been previously evaluated against human annotations in their original works, we test their utility in our simulated unreliable data environments by analyzing the entropy of the resulting importance scores across sentences drawn from data sources with varying levels of pollution.
As illustrated in \autoref{fig: entropy}, both MC-Shapley and RORA exhibit a consistent decrease in score entropy as the data pollution level increases. This is because token-level pollution progressively replaces ground-truth tokens with incorrect ones, leading to uniformly low scores across sentences, thereby reducing entropy.
Importantly, this pattern reflects the methods' sensitivity to data quality degradation, confirming their viability as robust indicators for identifying unreliable information for our \name{} system. We further provide a real example in \autoref{table: heatmap example} showing how the two methods capture the relative importance of each sentence.

\subsection{System Cost Evaluation}
\label{subsec: system cost}

\begin{table}[ht]
\small
\centering
\begin{tabular}{ccc}
\toprule
$n$ & \multicolumn{1}{c}{Total Gas Used (Cost)} & \multicolumn{1}{c}{Per-Update Gas (Cost)} \\
\midrule
1  & 71,277 (\$0.0258) & 71,277 (\$0.0258) \\
2  & 96,352 (\$0.0349) & 48,176 (\$0.0174) \\
5  & 211,492 (\$0.0766) & 42,298 (\$0.0153) \\
10 & 376,899 (\$0.1364) & 37,690 (\$0.0136) \\
15 & 502,565 (\$0.1820) & 33,504 (\$0.0121) \\
20 & 628,048 (\$0.2274) & 31,402 (\$0.0114) \\
\bottomrule
\end{tabular}
\caption{Gas usage and USD cost for providing feedback and updating the reliability scores at 0.09 gwei/gas and ETH = \$4,022.14 (on Oct 29, 2025). $n$ represents the number of data sources to update in the feedback. This mainly reflects the cost of the signature. Actual gas consumption also depends on the query length and other factors.}
\label{table: gas-costs}
\end{table}

Utilizing decentralized infrastructure like a public blockchain is typically not free, as it consumes the computational power of the network.
In the Ethereum network, \textit{gas} is the unit that measures the computational effort required to execute operations, like running a smart contract. Each operation incurs a fixed amount of gas, and users pay a fee based on this total, which compensates network participants for the resources they use. \autoref{table: gas-costs} provides the cost of the most important part of making \name a trustworthy infrastructure, namely, on-chain verification of feedback and score updates. As the number of score updates $n$ increases (i.e., the number of queries the \name{} has seen), the per-update cost decreases, reflecting approximately 55.8\% of gas efficiency gains from our batching operations, dropping from 2.6 cents per update (n=1) to 1.1 cents per update (n=20). This demonstrates the potential for \name to utilize batched queries and feedback to update the reliability evaluation while maintaining a reasonable cost range.

\section{Conclusion and Future Work}
We built \name, a decentralized RAG system that addresses data reliability challenges in real-world settings. Our \name{} dynamically computed and updated reliability scores for each data source, securing these scores on a blockchain to ensure both quality control and trust among data source owners. Through controlled experiments, we demonstrated that \name{} consistently outperforms centralized systems in unreliable data environments across different models, data pollution strategies, and reliability scoring methods. Notably, our system can even achieve performance comparable to that obtained in fully reliable environments. To the best of our knowledge, \name{} is the first blockchain-based decentralized RAG system that incorporates quality control mechanisms to handle noisy, real world-like data. We hope this work opens new research directions at the intersection of decentralized systems, RAG, and trustworthy AI.

While \name{} performs effectively when data sources provide full coverage for user queries (i.e., token-level pollution), it struggles under more challenging conditions. For example, when pollution occurs at the document level, its performance falls considerably short of the upper bound achieved by centralized systems in fully reliable environments, as we briefly discussed in \Cref{subsec: ce vs de}. This gap arises from a fundamental challenge in the data distribution: each data source provides only partial documents rather than complete coverage of the query space, such that the most reliable source may not contain the most relevant documents for a given query.
To address this limitation, a straightforward solution is to make reliability scores work adaptively rather than through a single fusion at the reranking stage. For instance, the system could first retrieve documents by relevance, then use a learned gate (classifier) to determine whether to apply reliability-based reranking or preserve the original relevance ordering for each query. Such an adaptive, query-conditioned pipeline, paired with our iterative source reliability updates, may close the performance gap in challenging scenarios with fine-grained, non-overlapping data sources.

Another promising direction for future research is to analyze the convergence rate of reliability scores in \name{}. Specifically, it remains unclear how many queries are required for the \name{} system to reach its performance upper bound, and which factors (e.g., query diversity, number of data sources, data source quality, etc.) most significantly influence the convergence rate of these reliability scores. A systematic investigation of these convergence dynamics would provide valuable insights into the system's scalability and practical deployment. We leave this exploration to future work.

\section*{Acknowledgements}
This work was partially supported by NSF IIS-2119531, IIS-2137396, IIS-2142827, IIS-2234058, OAC-2312973, NASA ULI 80NSSC23M0058 and Open Philanthropy. We also appreciate the support from the Foundation Models and Applications Lab of Lucy Institute and ND-IBM Tech Ethics Lab.




\bibliography{reference}
\bibliographystyle{mlsys2025}

\appendix


\end{document}